\documentclass[preprint,5p,twocolumn]{elsarticle}

\usepackage{algpseudocode}
\usepackage{algorithm}
\usepackage[cmex10]{amsmath}

\journal{Arxiv}

\begin{document}

\begin{frontmatter}
\title{Real-time High Resolution Fusion of Depth Maps on GPU}
\author{Dmitry S. Trifonov \thanks{e-mail:slonegg@gmail.com}\\Artec Group Inc./Keldysh Institute of Applied Mathematics RAS}


\begin{abstract}
A system for live high quality surface reconstruction using a single moving depth camera on a commodity hardware is presented. High accuracy and real-time frame rate is achieved by utilizing graphics hardware computing capabilities via OpenCL\texttrademark~and by using sparse data structure for volumetric surface representation. Depth sensor pose is estimated by combining serial texture registration algorithm with iterative closest points algorithm (ICP) aligning obtained depth map to the estimated scene model. Aligned surface is then fused into the scene. Kalman filter is used to improve fusion quality. Truncated signed distance function (TSDF) stored as block-based sparse buffer is used to represent surface. Use of sparse data structure greatly increases accuracy of scanned surfaces and maximum scanning area. Traditional GPU implementation of volumetric rendering and fusion algorithms were modified to exploit sparsity to achieve desired performance. Incorporation of texture registration for sensor pose estimation and Kalman filter for measurement integration improved accuracy and robustness of scanning process.
\end{abstract}


\end{frontmatter}

\section{Introduction}
With the advent of general purpose GPU computing, surface reconstruction algorithms that were originally purely offline have become real-time. Release of cheap depth sensors like Microsoft Kinect or Asus Xtion have made reconstruction algorithms even more appealing.

GPU implementations still have several restrictions. Recent implementations used a dense volumetric representation of the scanned model to perform quick surface fusion. This representation requires huge amounts of memory. For example, storing four bytes for $640^3$ voxel field would require $1Gb$ of GPU memory. Typically this is the maximum contiguous memory block that can be allocated on modern GPU. Such resolution is enough for Kinect for scanning small room environments (Newcombe et al. suggested that their system can be used to conveniently scan rooms with volumes $\le 7m^3$~\cite{NewcombeIHMKDKSHF11}), but doesn't expose the full potential of more precise scanners.

\begin{figure}[ht]
  \centering
  \includegraphics[width=3.2in]{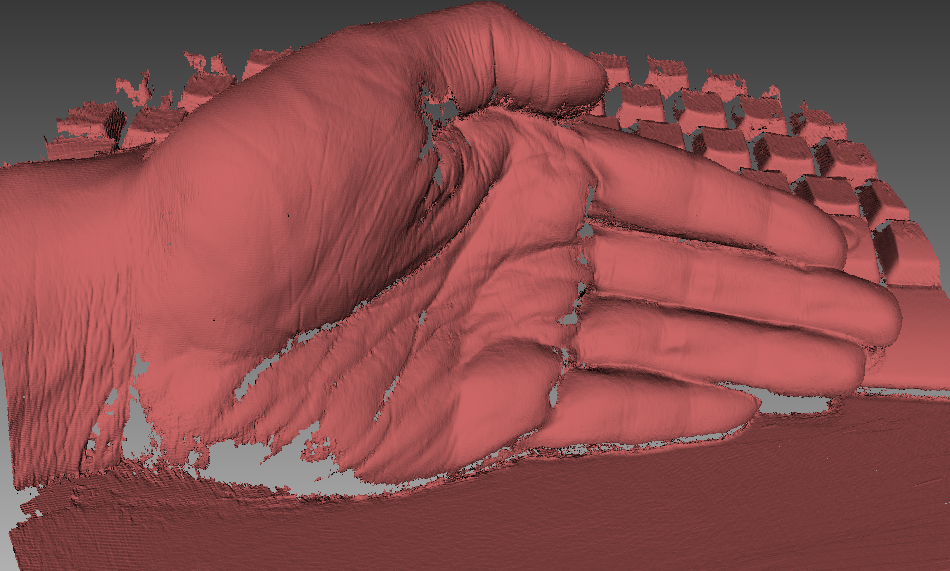}
  \caption{Scanning human hand with Artec\texttrademark~Spider scanner. Model is obtained in real-time using $4096^3$ voxel resolution. Size of the voxel is $0.15mm\times 0.15mm\times 0.15mm$.}
\end{figure}

To address this problem reconstruction algorithms were modified to use block-based sparse buffer for the volumetric model representation. This allows reconstruction of more detailed models with less memory footprint and higher performance.

Accurate optical scanners usually have small scanning range. Often not enough geometry features are captured within their working volume, thus use of only geometry information for surface pose estimation isn't always sufficient. In order to deal with this problem large scale GPU registration algorithm was combined with CPU texture registration to allow convenient scanning.

Additionally the variance in the input data is not uniform for the optical range scanners. Usually measurements are normally distributed with deviation growing quadratically with the scanned distance. Poor quality measurements may also appear near the edges of the scanned surface. In order to achieve best results all measurements have to be properly integrated accordingly to their relevance. To handle this problem a variant of the Kalman filter was implemented.

\begin{figure}[ht]
  \centering
  \includegraphics[height=2.6in]{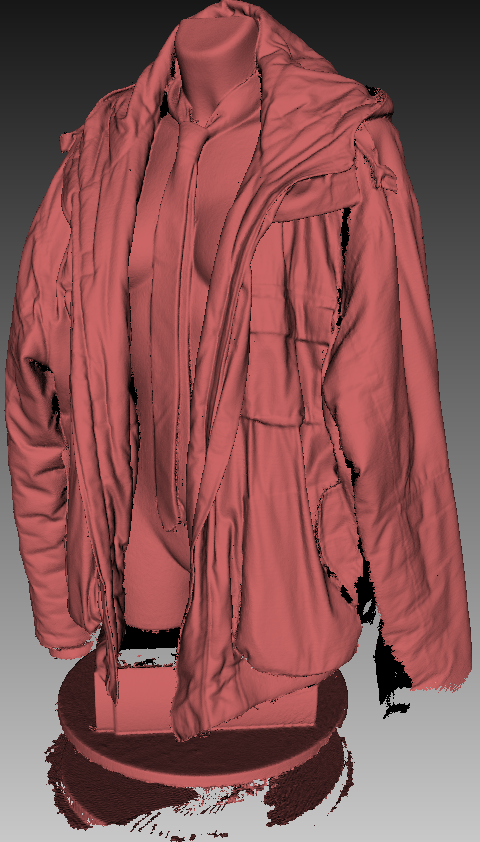}
  \includegraphics[height=2.6in]{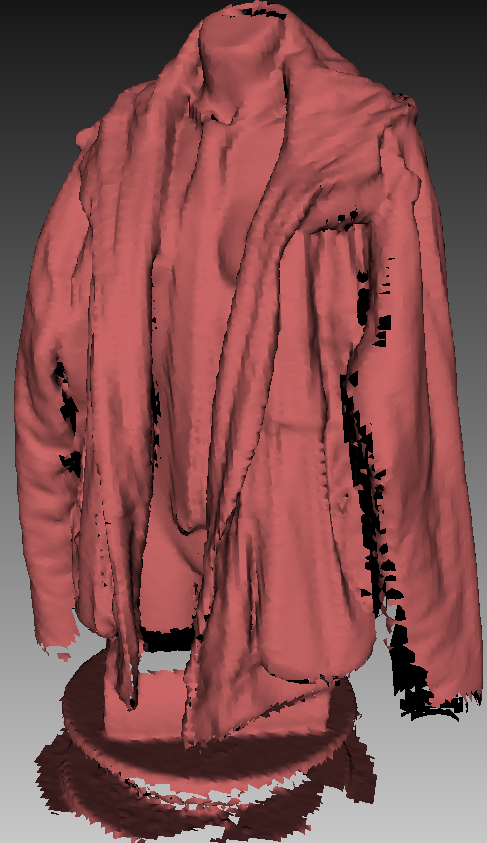}
  \caption{Scanning mannequin with Artec\texttrademark~Eva scanner. Left model is obtained in real-time using $3000^3$ voxel resolution and right is obtained with $500^3$ voxel resolution. Scanning box is $3m\times 3m\times 3m$.}
  \label{fig:scanning_comparison}
\end{figure}

\section{Related Work}

Interest in model acquisition has a long history. Many of the scanning researches share the same pipeline: surface acquisition, registration of the obtained surfaces and finally  fusion to obtain high quality model. In the earliest researches all of these stages were performed separately and offline~\cite{Curless_1996:VMB:237170.237269}. Thus interest was mostly focused in quality and stability improvement of each of this stages to allow finer model acquisition with less human intervention.

First real-time scanning applications used to store point cloud instead of fully polygonized model, this allowed to avoid costly fusion step and perform ICP registration as usual~\cite{Rusinkiewicz:2002:RMA:566654.566600}. During the scanning process some representation of the reconstructed model was rendered to the user, so it can control the scanning process. Then enough data was acquired user terminated the scanning process and offline global registration and fusion algorithms were performed to obtain high quality model. With such software almost anyone can digitize real-world objects.

With the rise of computation power real-time implementation of the costly fusion step become practicable~\cite{NewcombeIHMKDKSHF11}. Performing fusion in real-time decreases the time necessary to obtain high quality model, simplifies and pleases scanning process for the user, improves the registration quality.

Still scanning large areas or scanning with high resolution is an open issue. Several attempts were made to overcome memory footprint problem. Zeng et al. implemented fusion algorithm using octree-based volumetric surface representation~\cite{Zeng2013126}. Several other researchers implemented eviction of the unused data to the external storage and sliding of the scanning volume~\cite{MovingVolume}\cite{PCL}\cite{Whelan13icra}.

Eviction of the unused data to the external storage device seem to be very appealing approach for scanning large environments. But with dense scene representation very large amounts of data have to be streamed. For example, considering $512^3$ resolution with four bytes stored per voxel, with scanning volume of $1m \times 1m \times 1m$ and average moving speed of $0.5m/s$, streaming speed should be about $256Mb/s$. With the increase of resolution required streaming speed grows cubically, thus maximum moving speed quickly decreases and scanning process becomes uncomfortable. Other impediment is the performance, which also quickly degrades with the increase of resolution. Thus effective sparse representation is crucial for high resolution scanning.

\begin{figure}[ht]
  \centering
  \includegraphics[width=3.2in]{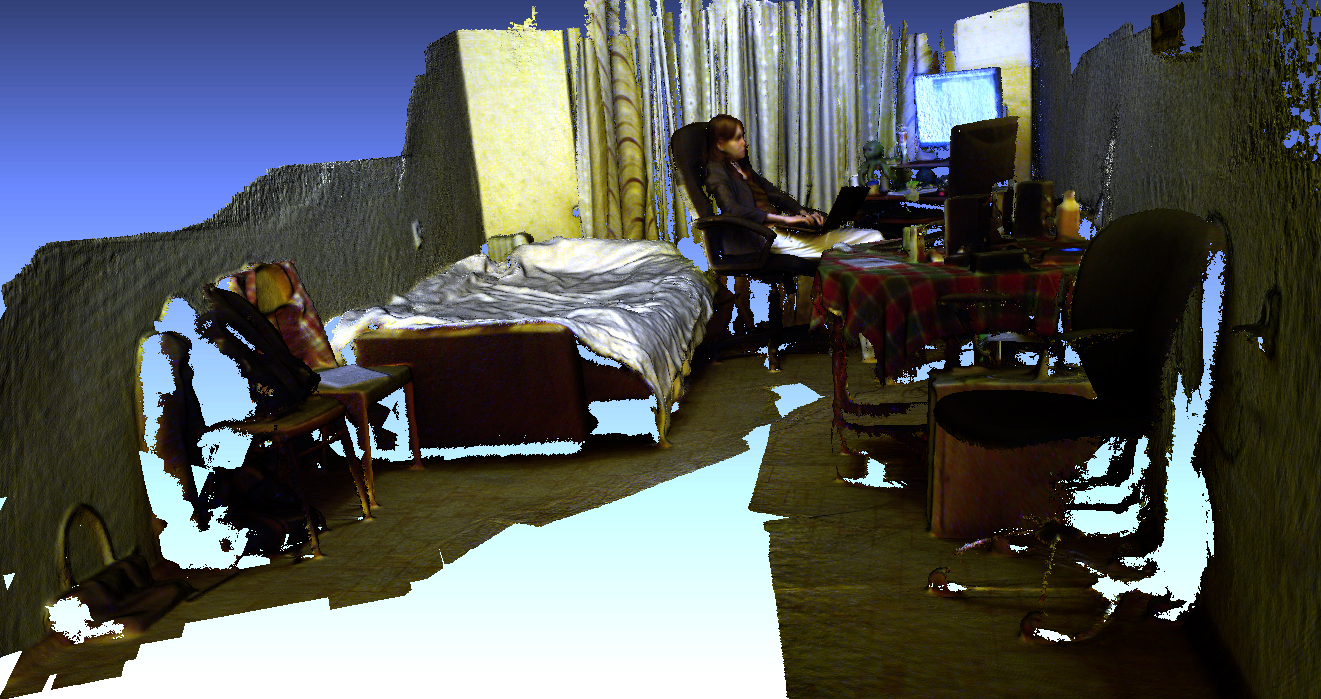}
  \caption{Room scanned in real-time with proposed method using Kinect\texttrademark~sensor with $4096^3$ voxel resolution. Texture mapped in post-processing.}
\end{figure}

\section{Pipeline Overview}

Fusion algorithm consists of several steps that are repeatedly executed during scanning.

\begin{enumerate}
  \item \textbf{Acquisition}. Obtain depth map from 3D scanner
  \item \textbf{Scene rendering}. Obtain depth and normal map of the currently reconstructed scene using previous camera position and orientation
  \item \textbf{Registration}. Estimate current sensor position and orientation by registering captured depth map to the scene depth map
  \item \textbf{Fusion}. Update scene using captured depth map, sensor position and orientation and possibly a weight map
\end{enumerate}

Same pipeline and same algorithms are used for all tested scanners ranging from high range Kinect to precision scanners with depth accuracy of 50 microns.

\section{Volumetric Scene Representation}

Truncated signed distance function (TSDF) is used to represent scene. Positive distance to the surface is stored in the voxel if it is placed outside surface and negative otherwise. If distance exceeds threshold $\delta$ by absolute value special mark $\chi$ is stored instead. Any algorithm processing voxel field, e.g. marching cubes, fusion or raycasting, encountering this value realizes that there is no isosurface passing nearby this voxel.

\begin{equation}
T(s,x)=
\left\{\begin{matrix}
\begin{aligned}
\rho (s,x)~&\text{if x is outside surface}\\ 
-\rho (s,x)~&\text{if x is inside surface}\\
\chi~&\text{if}~\rho (s,x)>\delta
\end{aligned}
\end{matrix}\right.
\end{equation}

Using TSDF instead of regular signed distance function with small $\delta$ allows to use only a few bits to represent relatively accurate models, thus decreasing memory consumption which is crucial in this application. In the proposed implementation eight bits are used to store distance. Use of TSDF also simplifies fusion algorithm, because only voxels near the surface have to be updated.

\begin{figure}[ht]
  \centering
  \includegraphics[width=3.2in]{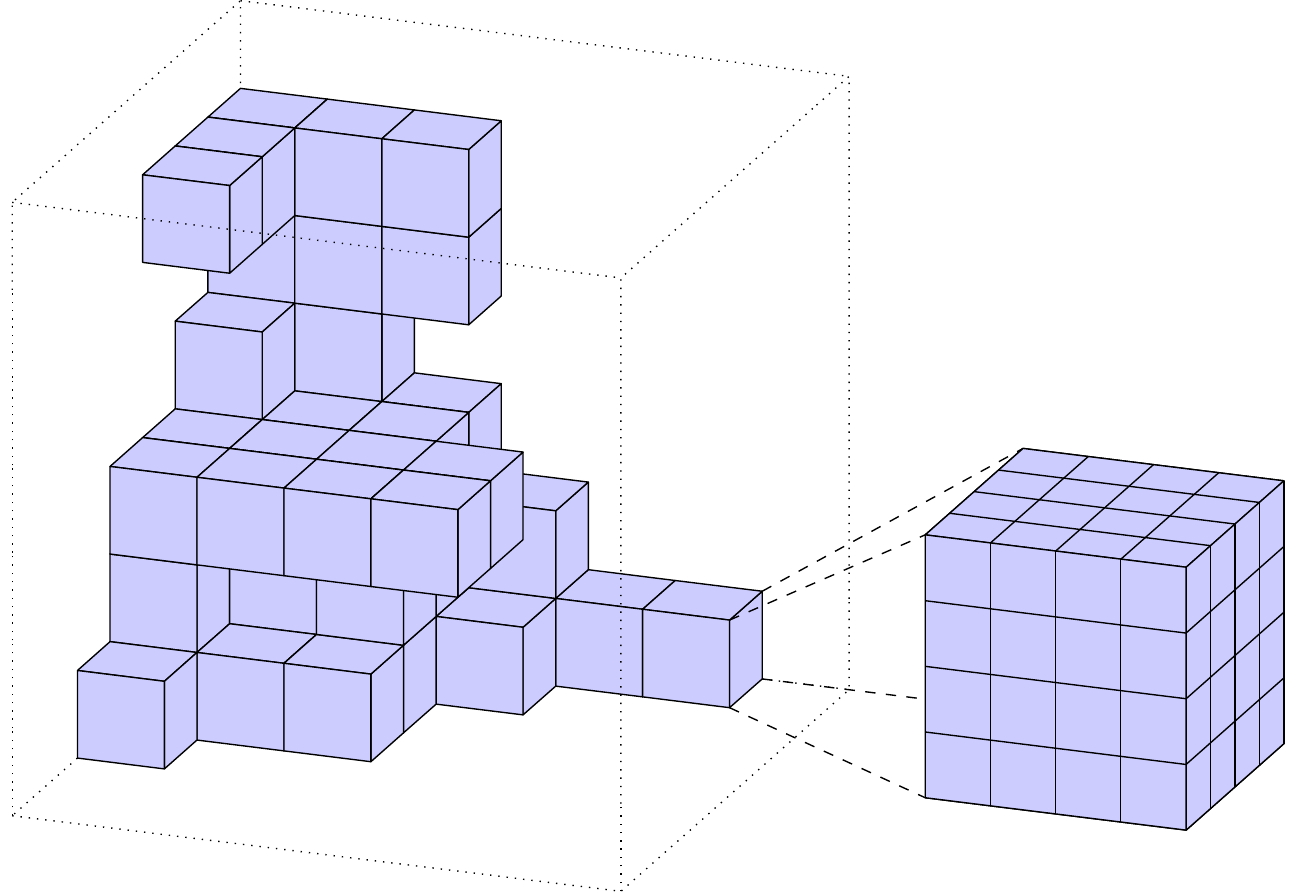}
  \caption{Block based sparse buffer storage of voxel field. Voxel field is subdivided into $N^3$ blocks. Each block is further subdivided into $M^3$ voxels. If anything except $\chi$ is going to be stored into voxel, then memory is allocated for the block containing it and filled with the necessary TSDF values. }
\end{figure}

\begin{figure}[ht]
  \centering
  \includegraphics[width=3.2in]{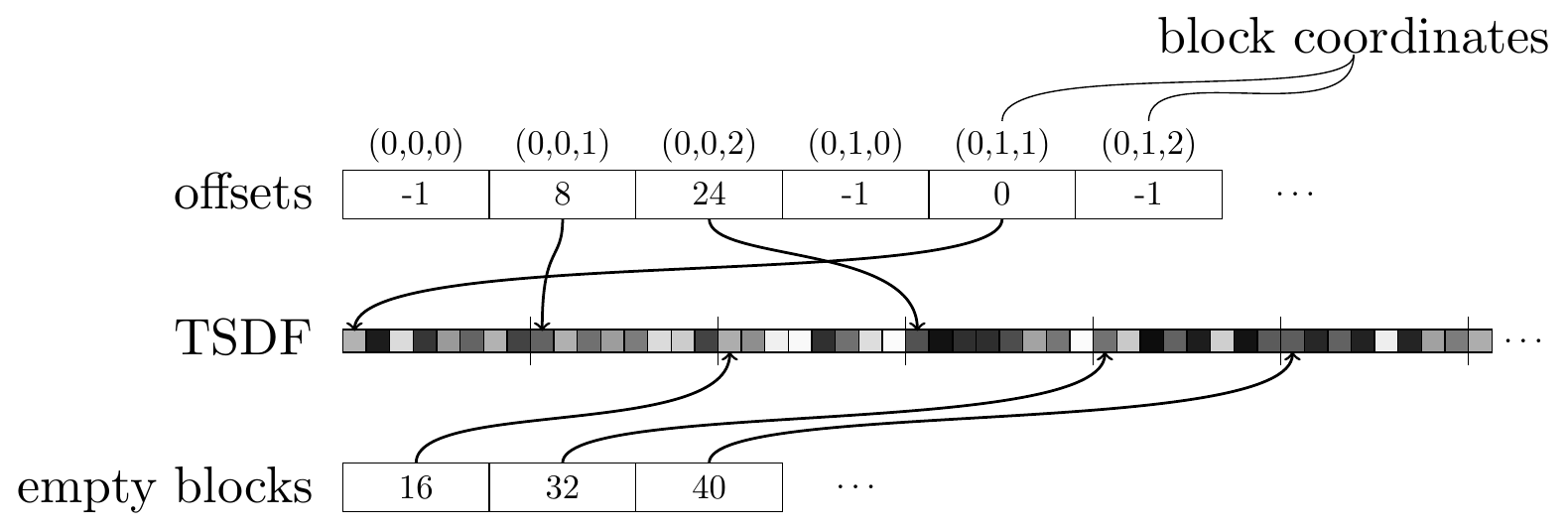}
  \caption{Data structure for storing sparse voxel field on GPU. Considering $N=3$ and $M=2$. TSDF values are color coded.}
  \label{fig:voxel_field_storage}
\end{figure}

TSDF is stored using block-based sparse buffer. It can be thought of as a two-level-deep hierarchical data structure where the domain of the buffer is subdivided into coarse $N^3$ blocks, each of them is either empty (i.e. TSDF value is $\chi$ inside it) and do not consume any space either is further subdivided into $M^3$ voxels. For simplicity assuming that resolution in each dimension is the same. At the GPU side two buffers are stored: one of them stores actual TSDF data and the other one stores a value for each coarse block, it can be either offset in the TSDF buffer, either -1 indicating that block is empty (fig.~\ref{fig:voxel_field_storage}). To get actual value of the TSDF two samples are necessary: offset in the second buffer and actual value of the function in the buffer with TSDF data.

Both buffers are preallocated on the GPU. Allocation and deallocation of blocks is done on the CPU by simply writing value to the specified element of the offset buffer. To do this list of empty blocks is maintained.

\subsection{Resolution and Memory Consumption}

Resolution is controlled by two parameters: number of blocks per axis - $N$, number of voxels per block axis - $M$. They cannot be varied arbitrary. Lower values of $M$ decrease performance, higher values increase memory consumption. Value of $N$ is limited by total available GPU memory. If $N$ is chosen very large, blocks are more dense, and more of them are required to reconstruct the same model. Eventually if resolution is too high algorithm will ran out of memory before scene is fully reconstructed. Thus, number of blocks per volume and number of voxels per block have to be carefully chosen depending on the characteristics of scanned objects.

\begin{figure}[ht]
  \centering
  \includegraphics[width=3in]{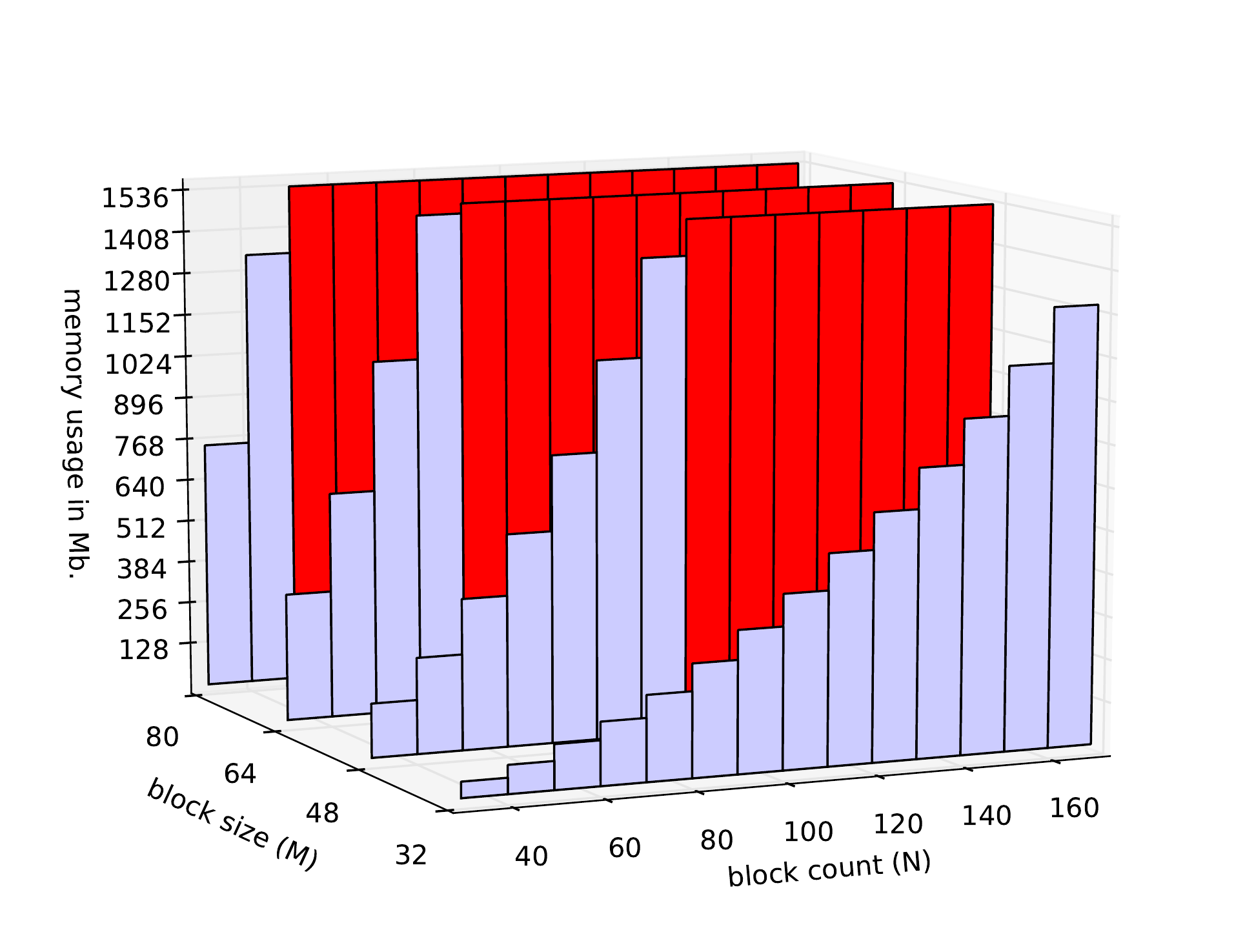}
  \caption{Memory consumption on test scene (fig.~\ref{fig:test_scene}). Size of the fusion box is $4m\times 4m\times 4m$. Using two bytes per voxel and four bytes to store offset. Thus total amount of memory in bytes is then $2*\langle\text{number of intersected blocks}\rangle* M^3 + 4 * N^3$. Drawing in red bars violating the memory threshold of 1536Mb.}
  \label{fig:memory_consumption_on_test_scene}
\end{figure}

One way two select $M$ and $N$ is to trial different values on a typical objects for scanning and select ones giving maximum resolution while simultaneously not violating desired memory threshold (fig.~\ref{fig:memory_consumption_on_test_scene}). 

Experiments show that on commodity hardware resolution up to $5000^3$ voxels is achievable. Presuming that maximum resolution of Kinect scanner is approximately 2mm~\cite{Khoshelham2012} described sparse data structure can be used to scan relatively large scenes with axial dimensions up to $10m$ with maximum resolution.

\begin{figure}[ht]
  \centering
  \includegraphics[width=3.2in]{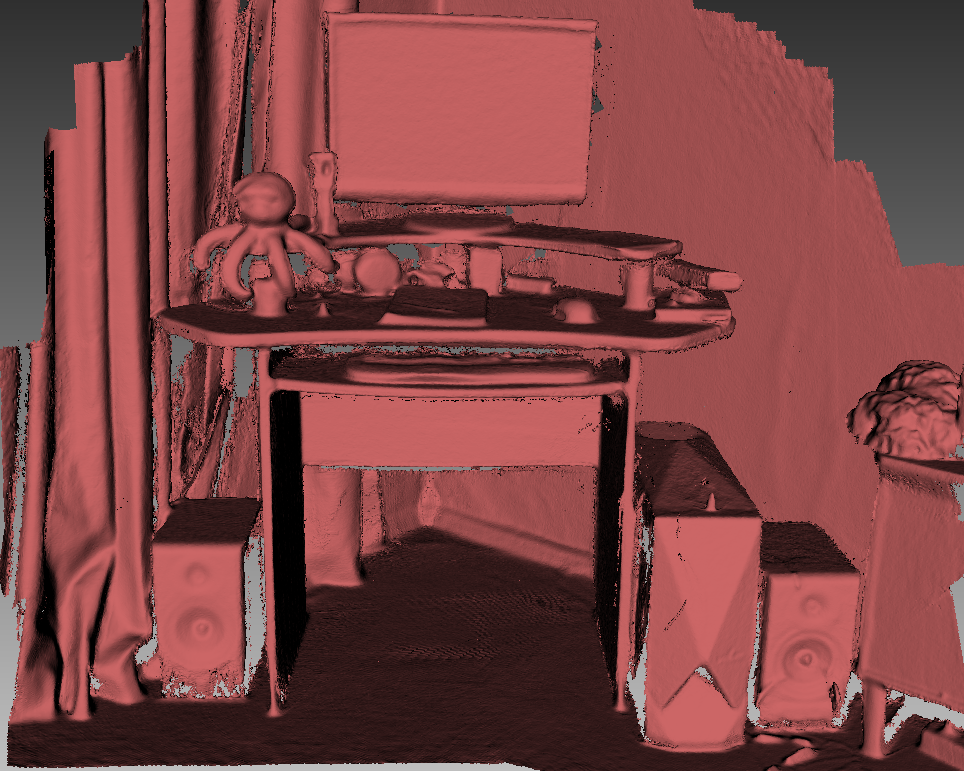}
  \caption{Test scene used in paper. Acquired using ASUS Xtion Pro. Resulting mesh is obtained using proposed method with $4m\times 4m\times 4m$ scanning box and $1024^3$ voxel grid.}
  \label{fig:test_scene}
\end{figure}

\section{Scene Rendering}

To display intermediate results to the user and to obtain depth and normal map for surface registration scene have to be rendered. Two approaches were tried to achieve the goal: marching cubes polygonization~\cite{Lorensen:1987:MCH:37401.37422},~\cite{Newman2006854} and raycasting. Marching cubes algorithm can perfectly exploit sparsity. Algorithm simply polygonizes non empty blocks in frustum, polygonized surfaces are then rendered to obtain depth and normal map. The implementation used is similar to one in NVIDIA CUDA SDK sample~\cite{NVSDK}.

This approach consists of three stages. During first stage occupied voxels and number of vertices produced by each voxel are calculated. Then prefix sum is used to from array of offsets in vertex array and array of coordinates of the occupied voxels. After that costly polygonization step can be performed only for occupied voxels. In order to reduce number of synchronizations between CPU and GPU occupied blocks are grouped together in batches. Size of the batch depends on the size of voxel block and allowed memory consumption. Above operations and rendering are performed for each batch.

Raycasting algorithm requires more elaborated approach. Performance of non modified version of the algorithm on high resolution sparse voxel field is far from satisfactory to scan in real-time. To exploit sparsity two additional helper textures are used. Each pixel of first one contains starting point for the raycasting and second one contains ending point. To fill this textures all occupied blocks are rendered with usual depth test and back face culling for first texture and with greater depth test and front face culling for second texture. Also depth values are clamped to near clip plane for the first texture and to far for the second.

\begin{figure}[ht]
  \centering
  \includegraphics[width=2.0in]{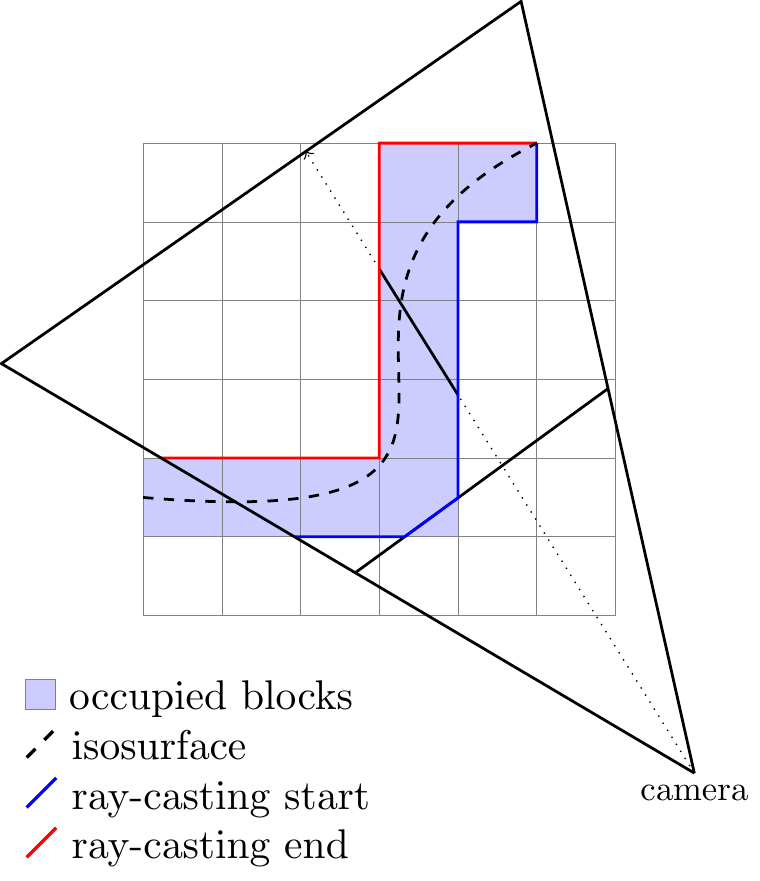}
  \caption{Reducing ray-casting range. Instead of traversing whole frustum volume only volume between front faces and back faces of occupied blocks is traversed.}
\end{figure}

These textures drastically shrink the volume need to be traversed by the raycasting algorithm. In proposed implementation raycasting algorithm is also split into two stages: coarse intersection is found during first stage, fine intersection and normal is found during second stage. This allows to achieve better occupancy because of reduced register usage in each stage.

\begin{figure}[ht]
  \centering
  \includegraphics[width=3.0in]{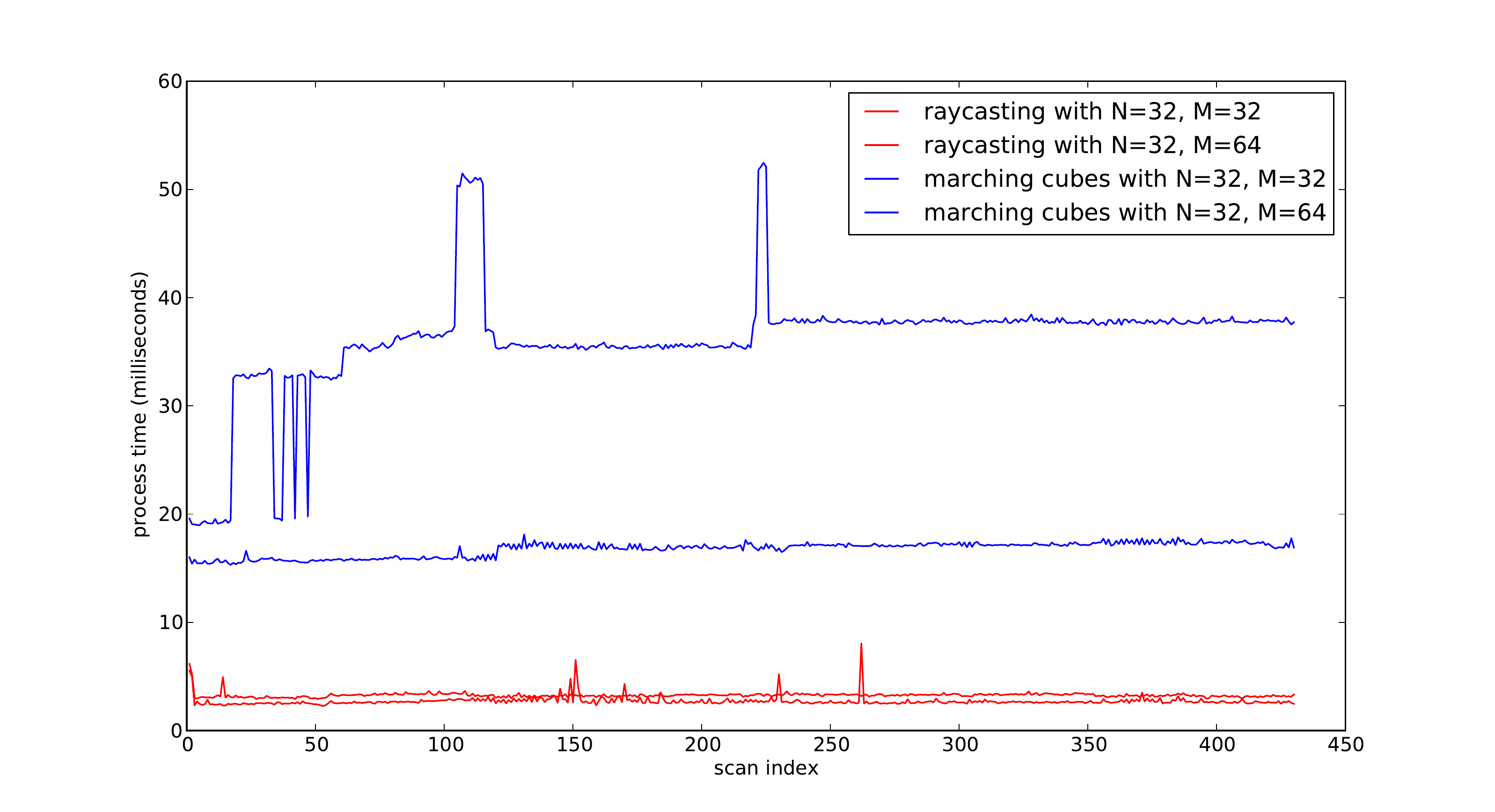}
  \caption{Comparison of the performance of the raycasting and marching cubes rendering on test scene (fig.~\ref{fig:test_scene}). Side of scanning volume cube: $6m$. Output size: 640x480. NVIDIA GeForce 460SE}
  \label{fig:rc_vs_mc}
\end{figure}

Marching cubes algorithm better exploit sparsity and almost doesn't degrade with increase of rendering resolution. On the other hand raycasting is generally faster and slowly degrades with increase of voxel field resolution and number of nonempty blocks. Performance tests show that for scanning application, especially with high range scanners, raycasting performs much better (fig.~\ref{fig:rc_vs_mc}).

\section{Registration}
To register obtained surface to the current estimation of the scene Chen-Medioni point-to-plane ICP framework with simple projective point matching and rejection based on distance and normal deviation is used~\cite{Chen:1992:OMR:138628.138633}\cite{Rusinkiewicz01efficientvariants}.

Once a collection of matched point pairs $(p_i,q_i)$ with normals $n_i$ is obtained, minimization of the following functional can be used to determine optimal rotation matrix $R$ and translation $t$ to bring obtained surface into alignment with scene.
\begin{equation}
\sum_{i}^N[(Rp_i+t-q_i)\cdot n_i] \rightarrow \min
\end{equation}
$R$ can be represented by series of rotations by $\alpha,\beta,\gamma$ around the $x,y,z$ axes respectively. By taking small angles assumption it can be represented as:
\begin{equation}
R=
\begin{bmatrix}
1&-\gamma&\beta\\
\gamma&1&-\alpha\\
-\beta&\alpha&1
\end{bmatrix}
\end{equation}
Minimization can be done by taking partial derivatives of functional with respect to $\alpha,\beta,\gamma,t_x,t_y,t_z$ and setting them to zero:
\begin{equation}
\label{eq:p2plane-minimization}
\begin{aligned}
&\sum_{i}^N
\begin{bmatrix}
c_ic_i^T & c_in_i^T\\ 
n_ic_i^T & n_in_i^T
\end{bmatrix}
\begin{bmatrix}
r\\
t
\end{bmatrix}
=
-\sum_{i}^N
\begin{bmatrix}
c_i(p_i-q_i)\cdot n_i\\
n_i(p_i-q_i)\cdot n_i
\end{bmatrix}\\
&c_i=p_i \times n_i\\
&r=\begin{bmatrix} \alpha & \beta& \gamma\end{bmatrix}^T\\
\end{aligned}
\end{equation}
Six dimensional equation is obtained. This can be easily solved on CPU using LU factorization. But to be able to register surfaces in real-time using all surface points fast reduction on GPU \cite{Reduction} have to be performed to obtain this equation.

In the proposed implementation two textures are used to store depth and normals, points $p_i,q_i$ are reconstructed using depth values. Matching, rejection and initial reduction step is done in a single OpenCL\texttrademark~kernel, second and further reduction stages are done in an additional kernel.

\subsection{Cooperation with CPU Registration}

Often not enough geometry information is captured with accurate low range scanner. One way to overcome this problem is to incorporate texture information (e.g. use an RGB-D SLAM method~\cite{Endres}), markers or another robust CPU registration algorithm to obtain initial transformation for the ICP.

In the proposed implementation initial transformation for the ICP is obtained by performing registration of the captured color frame to the previous frame. BRIEF features~\cite{Calonder12} are extracted from images and brought to alignment by performing RANSAC~\cite{Fischler:1981}. This helps, but equation (\ref{eq:p2plane-minimization}) is unstable whether not enough geometry information is present, thus small noise may spoil registration.

Transformations along eigenvectors corresponding small eigenvalues of covariance matrix of equation (\ref{eq:p2plane-minimization}) are unstable. To improve registration these transformations can be simply ignored.

To use simple thresholding scheme it is desirable to bring eigenvalues of (\ref{eq:p2plane-minimization}) to the same scale. This can be done by shrinking the box covering points $p_i,q_i$ to unit cube. The following transformation can be used to achieve this ($m$ is the center of the initial box, $S$ is a diagonal matrix with box dimensions):

\begin{equation}
\label{eq:shrink_coordinates}
\begin{aligned}
\hat{p}_i&=S^{-1}(p_i-m)\\
\hat{q}_i&=S^{-1}(q_i-m)\\
\hat{c}_i&=S^{-1}(c_i-m\times n_i)
\end{aligned}
\end{equation}

Substitution of (\ref{eq:shrink_coordinates}) into (\ref{eq:p2plane-minimization}) gives slightly different matrix equation.

\begin{equation}
\label{eq:p2plane-minimization-shrinked}
\begin{aligned}
&\sum_{i}^N
\begin{bmatrix}
\hat{c}_i\hat{c}_i^T & \hat{c}_in_i^T\\ 
n_i\hat{c}_i^T & n_in_i^T
\end{bmatrix}
\begin{bmatrix}
\hat{r}\\
\hat{t}
\end{bmatrix}
=
-\sum_{i}^N
\begin{bmatrix}
\hat{c}_iS(\hat{p}_i-\hat{q}_i)\cdot n_i\\
n_iS(\hat{p}_i-\hat{q}_i)\cdot n_i
\end{bmatrix}\\
&r=S\hat{r}\\
&t=\hat{t}+r\times m
\end{aligned}
\end{equation}

Initial rotation and translation can be found from modified ones using the relation:
\begin{equation}
\begin{aligned}
r&=S\hat{r}\\
t&=\hat{t}+r\times m
\end{aligned}
\end{equation}

New problem (\ref{eq:p2plane-minimization-shrinked}) is better in terms of accuracy, because summands are on the same scale now, and stability, because eigenvalues are also closer to each other. Now only underlying geometry and number of point pairs influence eigenvalues. Thus it seems reasonable to scale eigenvalues or threshold accordingly to the number of matched points. Finally $\hat{x}'=\begin{bmatrix}\hat{r}&\hat{t}\end{bmatrix}^T$ is found from $\hat{x}$ - solution of (\ref{eq:p2plane-minimization-shrinked}) as follows:
\begin{equation}
\begin{aligned}
&C=\sum_{i}^N
\begin{bmatrix}
\hat{c}_i\hat{c}_i^T & \hat{c}_in_i^T\\ 
n_i\hat{c}_i^T & n_in_i^T
\end{bmatrix}\\
&\lambda_1\hdots\lambda_6\text{ - eigenvalues of C}\\
&v_1\hdots v_6\text{ - eigenvectors of C}\\
&\hat{x}'=
\sum_{i=1}^6
\left\{\begin{matrix}
\begin{aligned}
&v_i(v_i \cdot \hat{x})~\text{if}~\lambda_i/N>\theta\\ 
&0~\text{otherwise}
\end{aligned}
\end{matrix}\right.
\end{aligned}
\end{equation}

The threshold $\theta$ is chosen empirically by examining eigenvalues of  (\ref{eq:p2plane-minimization-shrinked}) on a different types of objects. In the proposed implementation value $\theta=0.005$ was used.

\begin{figure}[ht]
  \centering
  \includegraphics[width=0.9in]{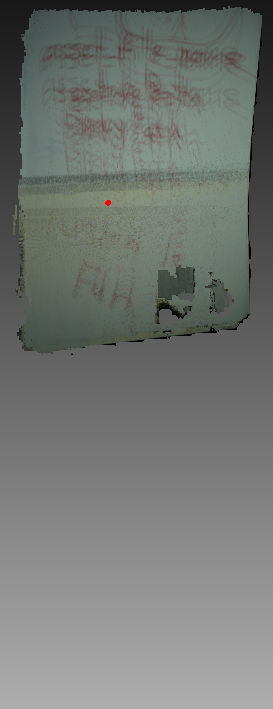}
  \includegraphics[width=0.9in]{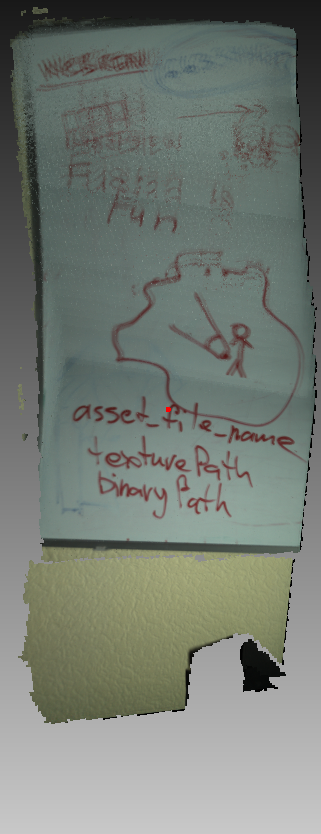}
  \includegraphics[width=0.9in]{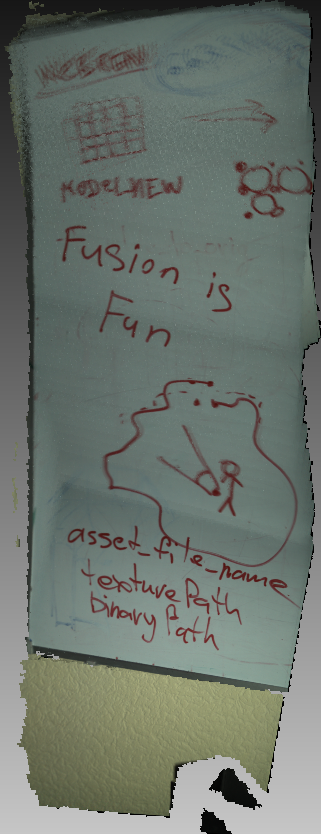}
  \caption{Comparison of registration methods on series of scans without geometric features. From left to right: serial ICP, serial ICP with texture, serial ICP with texture and eigenvalue analysis}
\end{figure}

\section{Fusion}

\subsection{Integration of Measurements}
Finally after surface obtained, preprocessed and registered it has to be merged with TSDF representing scene. Given two estimations $T$ - current estimation of scene TSDF and $T_k$ - surface measurement, $T$ can be updated by taking linear combination of two, but the care must be taken about $\chi$.

\begin{algorithm}
\caption{Scene TSDF Estimation}
\label{alg:scene_tsdf}
\begin{algorithmic}
\For {$\forall x \in dom(T_k) : T_k(x) \ne \chi$}
\If {$T(x) = \chi$}
	\State $T(x) \gets T_k(x)$
\Else
  	\State $T(x) \gets (1-w_k(x))T(x) + w_k(x) T_k(x)$
\EndIf
\If {$\left|T(x)\right| > \delta$}
	\State $T(x) \gets \chi$
\EndIf
\EndFor
\end{algorithmic}
\end{algorithm}

Here $w_k(x) \in (0,1)$ controls speed of update. It may be varied accordingly to the quality of the obtained data, i.e. it may be lowered for corner regions, because data there is usually more noisy.

By fixing $w_k(x)=N^{-1}$ and using the properties of normal deviates distribution of the final error can be estimated:

\begin{equation}
T_k(x)\in \mathcal{N}(0,\sigma^2) \Rightarrow 
T(x)=\sum_{k=1}^N\frac{T_k(x)}{N}\in\mathcal{N}\left(0,\frac{\sigma^2}{N}\right)
\end{equation}

Effectively this means that averaging of $N$ measurements decreases error $\sqrt{N}$ times. Values of $w_k~0.01$ give perfect quality, but user have to wait some time to accumulate enough measurements. One workaround is to use a weight function in addition to TSDF to store the sum of all weights for valid measurements \cite{NewcombeIHMKDKSHF11}:

\begin{equation}
W(x)=\sum_{k=1}^N
\left\{\begin{matrix}
\begin{aligned}
&w_k~\text{if}~T_k(x) \ne \chi\\ 
&0~\text{otherwise}
\end{aligned}
\end{matrix}\right.
\end{equation}

Algorithm~\ref{alg:scene_tsdf} has to be modified as follows:

\begin{algorithm}
\caption{Weighted Scene TSDF Estimation}
\label{alg:scene_tsdf_weight}
\begin{algorithmic}
\For {$\forall x \in dom(T_k) : T_k(x) \ne \chi$}
\If {$T(x) = \chi$}
	\State $T(x) \gets T_k(x)$
	\State $W(x) \gets w_k(x)$
\Else
  	\State $T(x) \gets \frac{W(x)T(x) + w_k(x) T_k(x)}{W(x)+w_k(x)}$
	\State $W(x) \gets \min(W(x) + w_k(x), W_{max})$
\EndIf
\If {$\left|T(x)\right| > \delta$}
	\State $T(x) \gets \chi$
\EndIf
\EndFor
\end{algorithmic}
\end{algorithm}

Specifying $W_{max}$ allows scanning dynamic environments, without this bound high values of $W(x)$ will discriminate new measurements $T_k(x)$. Although both algorithms eventually do exactly the same whether $W_{max}$ is reached, they both converge to the same value, weighted version faster converges in the beginning. Thus even if $w_k$ is chosen to be low, smooth surface is obtained faster. The cost is doubled memory consumption.

Both algorithms suffer from same drawbacks. First is that high amount of low quality data eventually worsens the approximation. This can happen for example if sensor is not moving much for a long period of time. Poor quality estimations at the model edges or at distance worsen the reconstruction quality. The other one is that it is impossible to achieve both quick response to the environment changes and good smoothing.

An attempt was made to overcome these problems by incorporating more sophisticated filtering scheme. A Kalman filter was chosen for its efficient recursive nature and property of giving statistically optimal estimate of the state variable~\cite{Welch:1995:IKF:897831}.

Because a very low amount of information can be stored per voxel the simplest filtering scheme is used. The value of the signed distance in a voxel $T(x)$ is treated as a state variable and $T_k(x)$ as a noisy measurement. Instead of assigning weights $w_k(x)$ to measurements their variances $p_k(x)$ are estimated and instead of storing weight sum $W(x)$ estimated process variance $P(x)$ is maintained for each voxel. For optical sensors like Kinect error deviation grows quadratically, thus $p_k(x)$ grows as the power of four. Process variance $Q$ is used to adjust response and smoothing.

\begin{algorithm}
\caption{Kalman Filter for TSDF Estimation}
\label{alg:scene_kalman}
\begin{algorithmic}
\For {$\forall x \in dom(T_k) : T_k(x) \ne \chi$}
\If {$T(x) = \chi$}
	\State $T(x) \gets T_k(x)$
	\State $P(x) \gets p_k(x)$
\Else
	\State $P(x) \gets P(x) + Q$\Comment{Predict}
	\State $K \gets P(x) / (P(x) + p_k(x))$
  	\State $T(x) \gets T(x) + K(T_k(x)-T(x))$\Comment{Correct}
	\State $P(x) \gets (1-K)P(x)$
\EndIf
\If {$\left|T(x)\right| > \delta$}
	\State $T(x) \gets \chi$
\EndIf
\EndFor
\end{algorithmic}
\end{algorithm}

\begin{figure}[ht]
  \centering
  \includegraphics[width=3.2in]{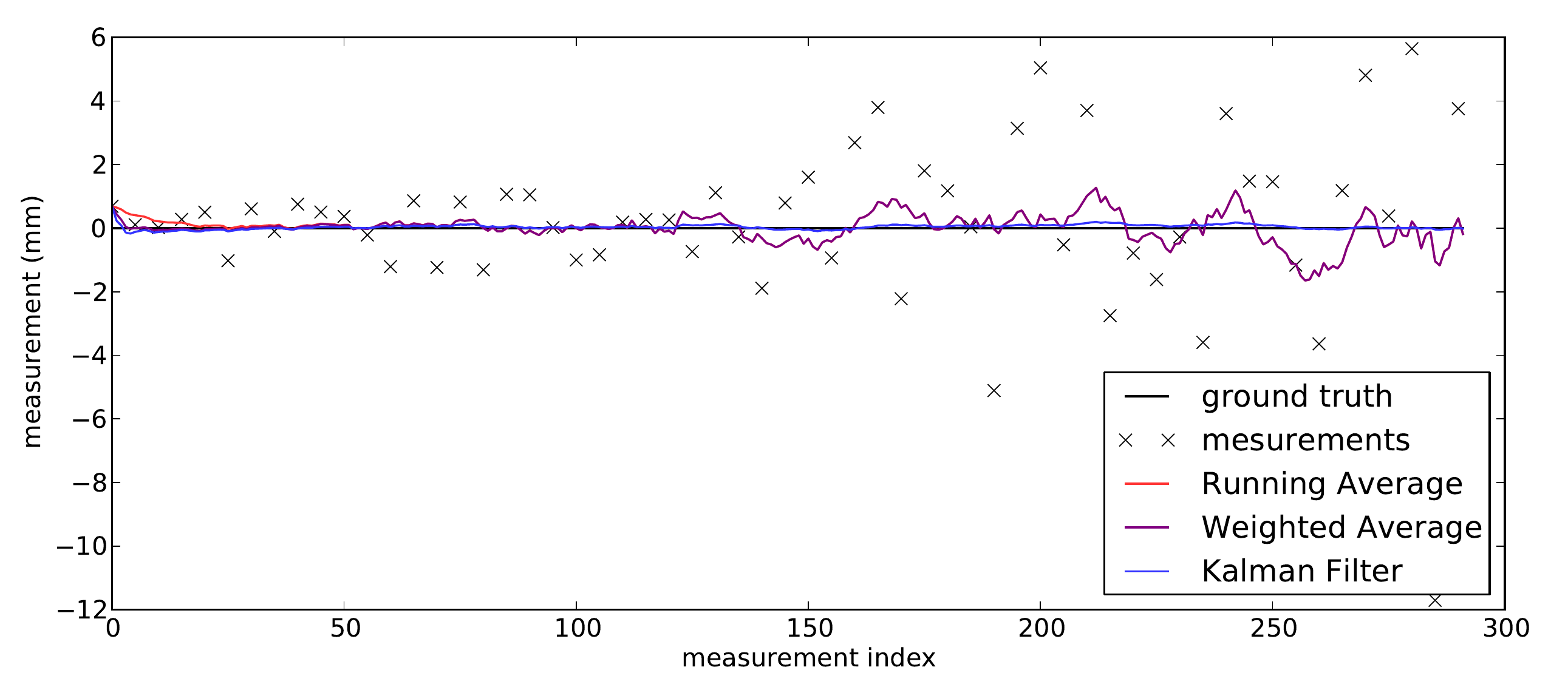}
  \caption{Comparison of measurement integration methods. Variance in the input data is continuously increasing. Parameters of the filters are chosen to give same smoothing. Parameters of the data distribution are chosen to resemble Kinect data at distance 1-2m.}
  \label{fig:measurements_with_increasing_variance}
\end{figure}

Kalman filters is good at protecting model from large amounts of poor data as compared to weighting methods. Increase of measurement variances lowers the gain $K$, thus poor data doesn't break the approximation of $T(x)$ and $P(x)$ (fig.~\ref{fig:measurements_with_increasing_variance}).

\begin{figure}[ht]
  \centering
  \includegraphics[width=3.2in]{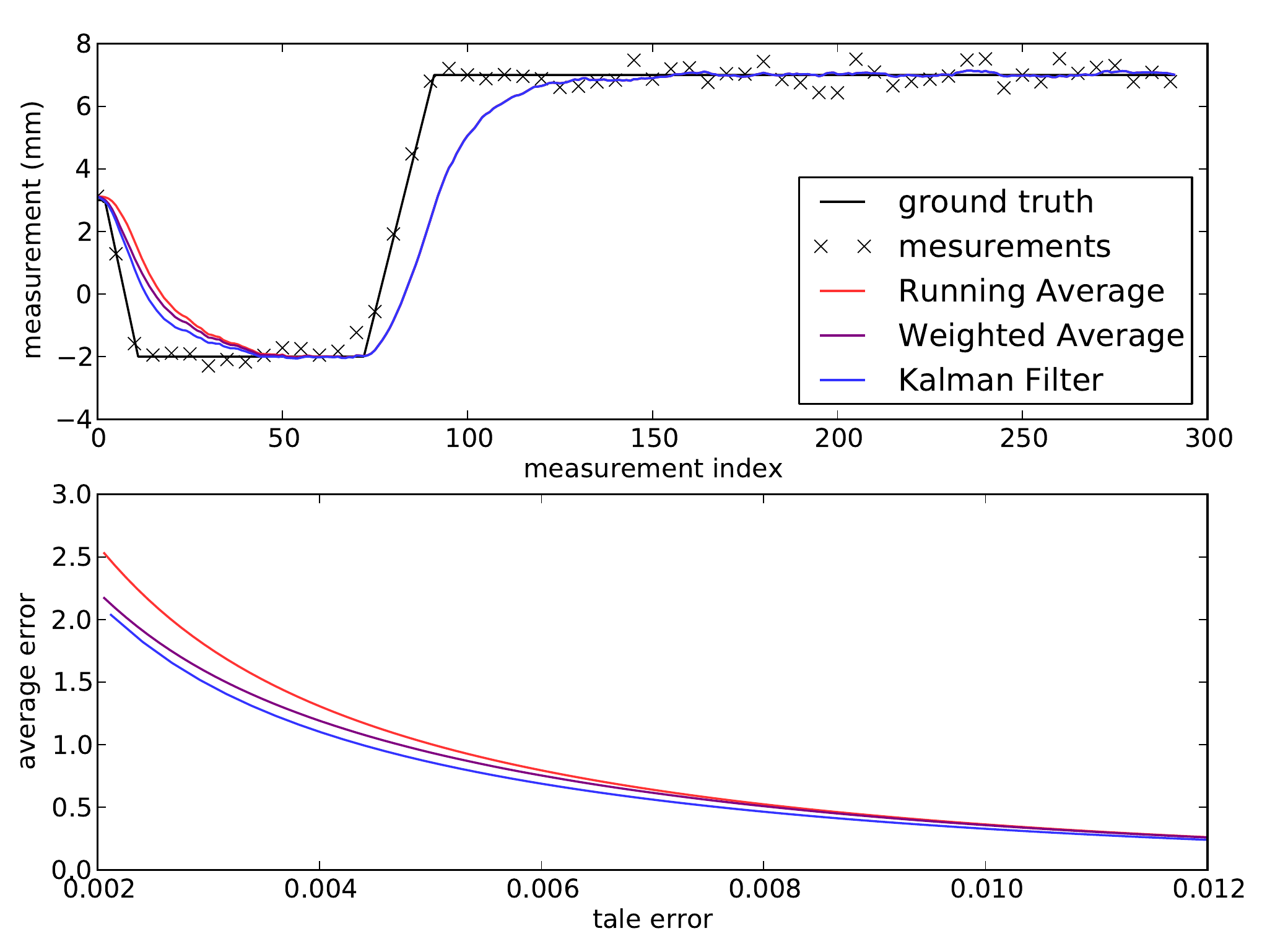}
  \caption{Comparison of measurement integration methods. On the bottom graphic sum of squared residuals for the whole history (average error) is compared with sum of squared residuals for the last 50 measurements (tale error). Parameters of the filters for the upper graphs are chosen to give the same tale error.}
  \label{fig:quality_versus_responce}
\end{figure}

Kalman filter handles quality versus response problem slightly better than weighting schemes (fig.~\ref{fig:quality_versus_responce}). Incorporation of more complex state dynamics and more state information may improve the situation, but in that case algorithm will consume much more memory. The reasonable direction of the research here seems to be in grouping filter parameters of nearby voxels.

In the described implementation $T_k(x)$ is not available explicitly, but it can be easily estimated for every $x$. The simplest way is to project $x$ onto the measured depth map and find distance between corresponding point in the depth map and $x$. Finer estimation may be obtained by performing several gradient steps in the depth map to find local minimum of the distance between surface and $x$. If no measurement is available for the specified $x$ then $\chi$ is simply taken as the value of $T_k(x)$

\subsection{Updating Sparse Structure}

\begin{figure}[ht]
  \centering
  \includegraphics[width=2in]{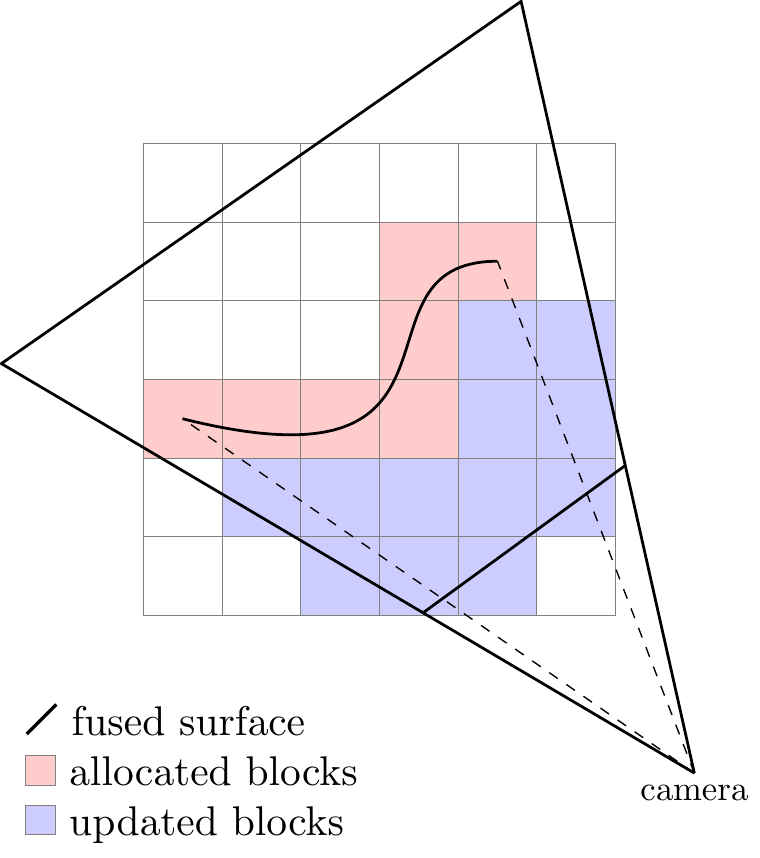}
  \caption{Choosing blocks for update during fusion step. Red blocks are going to be allocated if necessary and updated. Blue blocks are updated only if they are already allocated.}
  \label{fig:voxel_field_update}
\end{figure}

Domain of the $T_k(x)$ is defined as the set of voxel coordinates in the volume of all of the blocks visible from the current depth sensor position (located within the sensors frustum and not occluded by the measured surface). Although it is wasteful to allocate all these blocks. Only blocks intersected by the surface are allocated, other visible blocks are updated only if they were allocated during previous steps (fig.~\ref{fig:voxel_field_update}), thus $dom(T_k)$ is further reduced.

Visible blocks and blocks intersected by the surface can be found using occlusion queries. But with the resolution growth this approach scales poorly. A use of hierarchical structure like octree may help, but particularly in this specific problem a more simple approach may be exploited.

Because of fixed block size and resolution, coordinate of the grid block covering specified point can be simply computed. Simple OpenCL\texttrademark~kernel is adopted to sample the surface at some points and write block coordinates to the buffer, buffer is then read back to the CPU and necessary blocks are allocated. Sparsity of the sampling have to be chosen depending on the resolution of the voxel field to minimize amount of missed blocks. Though it doesn't matter that much if some blocks are missed, because scanner is continuously moving and missed blocks will be likely allocated next time.

\section{Conclusion}

\begin{figure}[ht]
  \centering
  \includegraphics[width=3.2in]{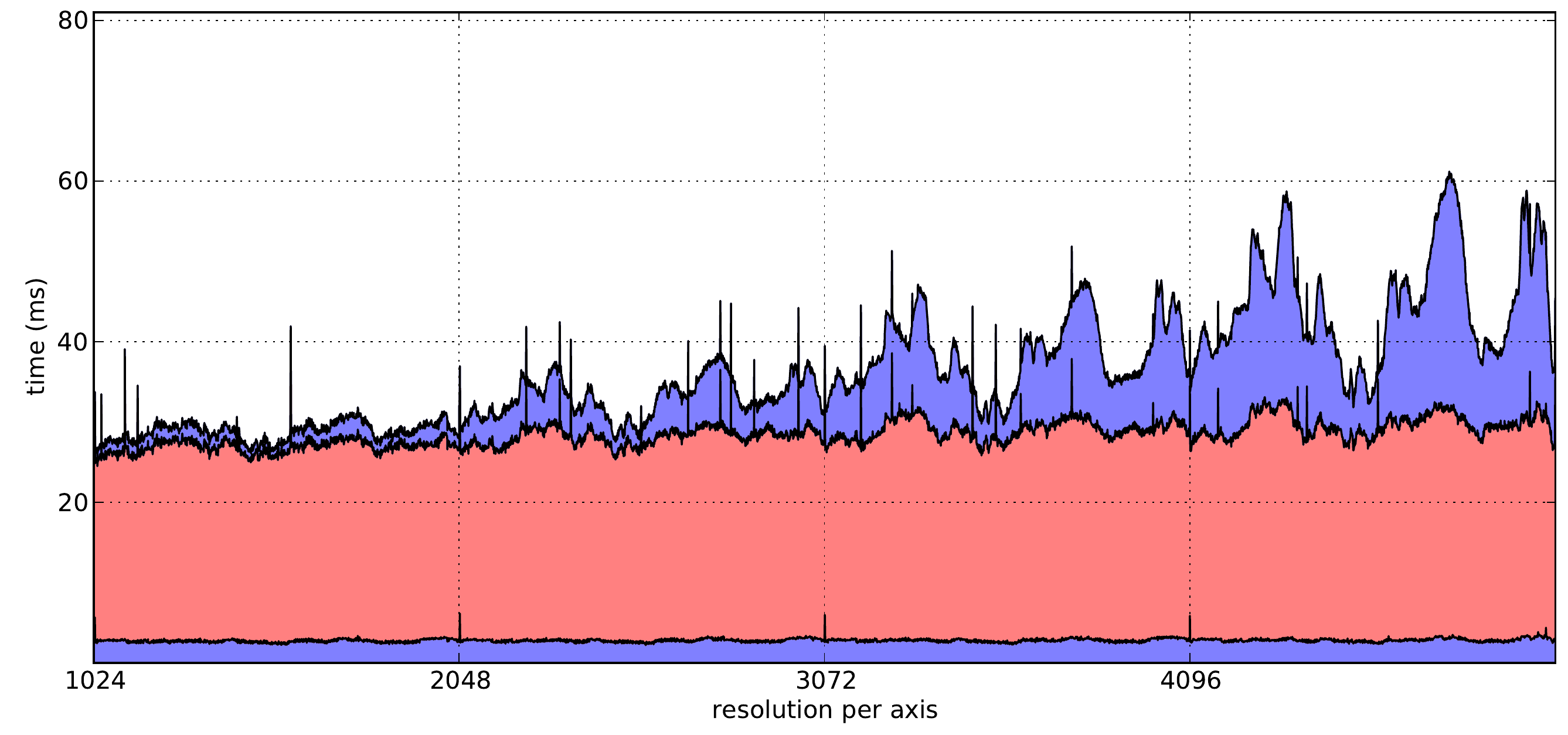}
  \caption{Performance of the proposed method on the test scene. Bottom to top: scene rendering via raycasting, ICP registration, measurement integration. NVIDIA GeForce 560 Ti.}
  \label{fig:overall_performance}
\end{figure}

Use of sparse data structure for real-time fusion improves significantly quality and maximum scanning volume without significant performance loss. Most algorithms have to be modified to exploit sparsity to achieve desired performance. Performance of proposed scheme scales well with the increase of resolution (fig.~\ref{fig:overall_performance}) or scanning area and the memory consumption is significantly reduced.

Still memory consumption is too high to digitize large outdoor environments or full interiors of the buildings. This problem can be resolved with the help of eviction schemes to exploit large volumes of HDD space or cloud storage. Implementation of the eviction should not be the problem with the described approach. Volume shift would require eviction of the TSDF data of the occupied blocks and rebuilding of the offset array. With current state of the hardware number of elements in the offset array should not exceed three millions in order to fit scene into available memory. Thus even naive rebuilding scheme  will give adequate performance.

Combination of ICP with texture registration greatly improves quality and robustness of the sensor pose estimation. This is especially noticeable while scanning with accurate low range scanner. Incorporation of Kalman filter for measurement integration protects reconstructed model from poor quality data, improves reconstruction quality. 

The utmost problem of the proposed method is the drift, occurring while scanning large areas due to local fashion of the registration. The method could be significantly improved by incorporation of global registration scheme. Other interesting directions of research include fusion with texture, non-rigid registration and fusion of deformable objects.

\section*{Acknowledgments}
Thanks to Pavel Rozanov and Anton Gudym for providing texture registration algorithm, to Yury Volodine for thorough review, to Hao Li for encouragement and valuable comments, to the whole Artec team for their support and knowledge.

\bibliographystyle{elsarticle-num}
\bibliography{bibliography}

\begin{thebibliography}{10}
\expandafter\ifx\csname url\endcsname\relax
  \def\url#1{\texttt{#1}}\fi
\expandafter\ifx\csname urlprefix\endcsname\relax\def\urlprefix{URL }\fi
\expandafter\ifx\csname href\endcsname\relax
  \def\href#1#2{#2} \def\path#1{#1}\fi

\bibitem{NewcombeIHMKDKSHF11}
R.~A. Newcombe, S.~Izadi, O.~Hilliges, D.~Molyneaux, D.~Kim, A.~J. Davison,
  P.~Kohli, J.~Shotton, S.~Hodges, A.~W. Fitzgibbon, Kinectfusion: Real-time
  dense surface mapping and tracking, in: ISMAR, 2011, pp. 127--136.

\bibitem{Curless_1996:VMB:237170.237269}
B.~Curless, M.~Levoy, A volumetric method for building complex models from
  range images, in: Proceedings of the 23rd annual conference on Computer
  graphics and interactive techniques, SIGGRAPH '96, ACM, New York, NY, USA,
  1996, pp. 303--312.
\newblock \href {http://dx.doi.org/10.1145/237170.237269}
  {\path{doi:10.1145/237170.237269}}.

\bibitem{Rusinkiewicz:2002:RMA:566654.566600}
S.~Rusinkiewicz, O.~Hall-Holt, M.~Levoy, Real-time 3d model acquisition, ACM
  Trans. Graph. 21~(3) (2002) 438--446.
\newblock \href {http://dx.doi.org/10.1145/566654.566600}
  {\path{doi:10.1145/566654.566600}}.

\bibitem{Zeng2013126}
M.~Zeng, F.~Zhao, J.~Zheng, X.~Liu, Octree-based fusion for realtime 3d
  reconstruction, Graphical Models 75~(3) (2013) 126 -- 136, computational
  Visual Media Conference 2012.
\newblock \href {http://dx.doi.org/10.1016/j.gmod.2012.09.002}
  {\path{doi:10.1016/j.gmod.2012.09.002}}.

\bibitem{MovingVolume}
H.~Roth, M.~Vona, Moving volume kinectfusion, in: Proceedings of the British
  Machine Vision Conference, BMVA Press, 2012, pp. 112.1--112.11.
\newblock \href {http://dx.doi.org/http://dx.doi.org/10.5244/C.26.112}
  {\path{doi:http://dx.doi.org/10.5244/C.26.112}}.

\bibitem{PCL}
F.~Heredia, R.~Favier, Kinfu Large Scale in PCL.\\
  \url{http://www.pointclouds.org/blog/srcs/} (2012).

\bibitem{Whelan13icra}
T.~Whelan, H.~Johannsson, M.~Kaess, J.~Leonard, J.~McDonald, Robust real-time
  visual odometry for dense {RGB-D} mapping, in: IEEE Intl. Conf. on Robotics
  and Automation, ICRA, Karlsruhe, Germany, 2013, to appear.

\bibitem{Khoshelham2012}
K.~Khoshelham, S.~O. Elberink, Accuracy and resolution of kinect depth data for
  indoor mapping applications, Sensors 12~(2) (2012) 1437--1454.
\newblock \href {http://dx.doi.org/10.3390/s120201437}
  {\path{doi:10.3390/s120201437}}.

\bibitem{Lorensen:1987:MCH:37401.37422}
W.~E. Lorensen, H.~E. Cline, Marching cubes: A high resolution 3d surface
  construction algorithm, in: Proceedings of the 14th annual conference on
  Computer graphics and interactive techniques, SIGGRAPH '87, ACM, New York,
  NY, USA, 1987, pp. 163--169.
\newblock \href {http://dx.doi.org/10.1145/37401.37422}
  {\path{doi:10.1145/37401.37422}}.

\bibitem{Newman2006854}
T.~S. Newman, H.~Yi, A survey of the marching cubes algorithm, Computers \&
  Graphics 30~(5) (2006) 854 -- 879.
\newblock \href {http://dx.doi.org/10.1016/j.cag.2006.07.021}
  {\path{doi:10.1016/j.cag.2006.07.021}}.

\bibitem{NVSDK}
NVIDIA, CUDA SDK, Marching Cubes Isosurfaces \\
  \url{http://docs.nvidia.com/cuda/cuda-samples/index.html} (2008).

\bibitem{Chen:1992:OMR:138628.138633}
Y.~Chen, G.~Medioni, Object modelling by registration of multiple range images,
  Image Vision Comput. 10~(3) (1992) 145--155.
\newblock \href {http://dx.doi.org/10.1016/0262-8856(92)90066-C}
  {\path{doi:10.1016/0262-8856(92)90066-C}}.

\bibitem{Rusinkiewicz01efficientvariants}
S.~Rusinkiewicz, M.~Levoy, Efficient variants of the icp algorithm, in:
  International Conference on 3-D Digital Imaging and Modelling, 2001.

\bibitem{Reduction}
B.~Catanzaro, Opencl™ optimization case study: Simple reductions, \\
  \url{http://developer.amd.com/resources/
  documentation-articles/articles-whitepapers/
  opencl-optimization-case-study-simple-reductions/} (2010).

\bibitem{Endres}
F.~Endres, J.~Hess, N.~Engelhard, J.~Sturm, D.~Cremers, W.~Burgard, An
  evaluation of the rgb-d slam system, in: Robotics and Automation (ICRA), 2012
  IEEE International Conference on, 2012, pp. 1691--1696.
\newblock \href {http://dx.doi.org/10.1109/ICRA.2012.6225199}
  {\path{doi:10.1109/ICRA.2012.6225199}}.

\bibitem{Calonder12}
M.~Calonder, V.~Lepetit, M.~Ozuysal, T.~Trzcinski, C.~Strecha, P.~Fua,
  {{BRIEF}: Computing a Local Binary Descriptor Very Fast}, IEEE Transactions
  on Pattern Analysis and Machine Intelligence 34~(7) (2012) 1281--1298.

\bibitem{Fischler:1981}
M.~A. Fischler, R.~C. Bolles,
  \href{http://doi.acm.org/10.1145/358669.358692}{Random sample consensus: a
  paradigm for model fitting with applications to image analysis and automated
  cartography}, Commun. ACM 24~(6) (1981) 381--395.
\newblock \href {http://dx.doi.org/10.1145/358669.358692}
  {\path{doi:10.1145/358669.358692}}.
\newline\urlprefix\url{http://doi.acm.org/10.1145/358669.358692}

\bibitem{Welch:1995:IKF:897831}
G.~Welch, G.~Bishop, An introduction to the kalman filter, Tech. rep., Chapel
  Hill, NC, USA (1995).

\end{thebibliography}

\end{document}